\documentclass[review]{agujournal2019} 
\usepackage{url} 
\usepackage{soul}
\usepackage[english]{babel}
\usepackage{float}
\usepackage{apacite}
\usepackage{multirow}
\usepackage{array}

\draftfalse
\journalname{Journal of Geophysical Research - Space Physics}
\newcommand{\degree}{$^\circ$}
\newcommand{\drag}{~m s$^{-1}$ day$^{-1}$}
\newcommand{\wind}{~m s$^{-1}$}

\newcommand{\Fig}{\textbf{Figure}~}
\newcommand{\Figs}{\textbf{Figures}~}

\usepackage{booktabs}

\usepackage{color}

\begin{document}

\title{Impact of Gravity Waves From Tropospheric and Non-tropospheric Sources on the  Middle and Upper Atmosphere and Comparison with ICON/MIGHTI Winds}

\authors{Erdal Yi\u git\affil{1,2}, Alexander S. Medvedev\affil{3}, Ayden L. S. Gann\affil{1}, Gary P. Klaassen\affil{4}, Douglas E. Rowland\affil{2}}

\affiliation{1}{George Mason University, Department of Physics and Astronomy, Space Weather Lab, Fairfax, VA, USA.}
\affiliation{2}{NASA Goddard Space Flight Center, ITM Lab (675), Greenbelt, MD, USA.}
\affiliation{3}{Max Planck Institute for Solar System Research, G\"ottingen, Germany.}
\affiliation{4}{Department of Earth and Space Science and Engineering, York University, Toronto, ON, Canada.}

\correspondingauthor{Erdal Yi\u git}{eyigit@gmu.edu (\today)}

\begin{keypoints}
\item  Gravity waves from tropospheric and extra-tropospheric sources are studied with CMAT2 and whole atmosphere gravity wave parameterization
\item  Waves from sources distributed over all heights produce the greatest dynamical and thermal effects in the thermosphere
\item  Comparison with ICON-MIGHTI observations shows CMAT2 reproduces the basic structure of thermospheric winds
\end{keypoints}

\begin{abstract}
We study the dynamical and thermal roles of  internal gravity waves generated in the troposphere and above using the Coupled Middle Atmosphere Thermosphere-2 General Circulation Model. This model incorporates the whole atmosphere nonlinear gravity wave parameterization and its extension to include non-tropospheric sources.  We conducted model experiments for northern summer solstice conditions, first including only tropospheric sources, then including sources localized at 50 and 90 km, and uniformly distributed over all heights. The simulated differences in mean temperature and horizontal winds demonstrate that gravity waves produce the greatest dynamical and thermal changes in the latter case compared to the localized sources. While the gravity wave drag is longitudinally uniform in the lower thermosphere, it is more localized in the upper thermosphere in all the simulations.  Waves from uniformly distributed sources  increase the longitudinal variability of zonal winds in the thermosphere up to $\sim$150 km. Gravity wave effects exhibit different local time variations in the lower thermosphere (100--140 km) than in the upper thermosphere. In the upper thermosphere, gravity wave effects are stronger during the day than at night. In contrast,  nighttime gravity wave effects are stronger than the daytime ones in the lower thermosphere. Finally, a comparison with ICON-MIGHTI observations shows that the model reproduces the basic structure of thermospheric winds, performing  better with zonal winds than with meridional winds. Adding non-tropospheric wave sources modifies wind structures in wave-breaking regions, but does not improve the global statistical comparison. 
\end{abstract}

\section*{Plain Language Summary}
Atmospheric gravity (buoyancy) waves are present in all planetary atmospheres. They alter the thermal and dynamical structure of Earth's atmosphere. Using a global-scale numerical model, we study the relative importance of primary (tropospheric) and non-tropospheric gravity waves. Global effects of these waves are largely captured by considering only  those that originate in the lower atmosphere.
However, the non-tropospheric sources contribute additional effects to the upper atmosphere, especially when they are assumed to be distributed over all heights. The importance of vertically localized sources decreases with the source height. A comparison of the model with ICON satellite observations demonstrates that the model can adequately reproduce the basic structure of thermospheric circulation when only tropospheric wave sources are included.  
%
%

\section{Introduction}
Internal gravity waves play a crucial role in the vertical coupling in the Earth's whole atmosphere system, from the ground to the exobase \cite{Fritts.Alexander2003_GravityWave, Yigit.Medvedev2015_InternalWave}. Their effects comprise acceleration/deceleration (or dynamical effects), heating/cooling (or thermal effects), and compositional mixing of the neutral gas \cite{Yigit.etal2009_ModelingEffects, Yigit.etal2012_DynamicalEffects, Yigit.etal2021_EffectsLatitudeDependent, Walterscheid.Hickey2012_GravityWave, Miyoshi.etal2014_GlobalView, Gavrilov.etal2020_ThermalEffects} and these effects strongly vary with altitude, season, and latitude. 
Gravity waves are ubiquitous in all stably stratified planetary atmospheres and ionospheres \cite{Yigit.Medvedev2019_ObscureWaves}. They have been routinely observed in Earth's atmosphere from the troposphere to the thermosphere during all seasons by airglow imagers, lidars, GPS radio occultation measurements, radars, and satellites \cite{Ern.etal2017_DirectionalGravity, Rapp.etal2018_IntercomparisonStratospheric, Stober.etal2023_IdentifyingGravity, Nayak.Yigit2019_VariationSmall, Garcia.etal2016_MediumscaleGravity, Trinh.etal2015_ComprehensiveObservational, Forbes.etal2016_GravityWaveinduced, England.etal2020_ObservationThermospheric, Nyassor.etal2022_InvestigationsConcentric, Conte.etal2022_ComparisonMLT, Nyassor.etal2022_SourcesConcentric, Huang.etal2021_InvestigationSpectral, Hertzog.etal2008_EstimationGravity, Nyassor.etal2025_MomentumFlux}.  One technical challenge is that there is no unique way to retrieve gravity wave properties from measurements. Depending on the techniques used, the observed  gravity wave characteristics may vary from method to method \cite{John.Kumar2013_DiscussionMethods, Sakib.Yigit2022_BriefOverview}.  Observations and models generally provide insight into different, but in principle complementary, aspects of gravity waves in the atmosphere. Namely, observations often quantify gravity wave activity in terms of wave potential energy per unit mass, vertical fluxes of horizontal momentum, amplitudes of temperature and density fluctuations. On the other hand, most global scale models seek to quantify the effects of gravity waves on  the mean flow by representing the contributions of small-scale gravity waves to the momentum, energy and mixing budget of the atmosphere \cite{Yigit.etal2021_EffectsLatitudeDependent, Yigit.etal2014_SimulatedVariability, Miyoshi.Yigit2019_ImpactGravity}.
Here we seek to employ the latter approach of parameterizing gravity wave effects from the tropopause to the upper thermosphere in a first-principles general circulation model (GCM) and comparing our simulated wind fields with observations from NASA's Ionospheric Connection Explorer (ICON) satellite. 

There are a number of sources of gravity waves  in Earth's atmosphere and ionosphere associated with topography, meteorological processes, and space weather. They can be located at all heights and produce harmonics with different characteristics. Since the amplitudes of gravity waves generally grow with altitude due to the decrease of air density, nonlinear processes become also more important. The harmonics generated by these weakly or strongly nonlinear processes acting on a propagating ``primary" wave are often referred to as ``secondary" harmonics.  Secondary (and higher-order) gravity waves can be generated in the middle atmosphere by the dissipating/breaking  primary harmonics propagating from the troposphere \cite{Holton.Alexander1999_GravityWaves}. Some of these waves can propagate further up into the thermosphere and contribute to the momentum budget there, as was suggested by previous theoretical modeling studies \cite{Vadas.etal2018_ExcitationSecondary}. Since the generation mechanism of such harmonics is not well understood and quantified,  the relative importance of the secondary waves is difficult to assess. The goal of this study is to narrow the knowledge gap by introducing proxies for unknown forcing mechanisms in the middle atmosphere, including secondary waves. 

It is often challenging to directly relate observed wave-like signatures to specific sources of gravity waves. While a number of studies interpreted observations of wave-like structures as secondary gravity waves \cite{deWit.etal2017_UnexpectedClimatological, Kogure.etal2020_FirstDirect}, some others could not clearly distinguish between the primary and higher-order  \cite{Vincent.etal2013_GravityWave}. On the other hand, numerical modeling provides a more practical framework to  separate these different sources from each other, since one can turn them on and off in  \textit{numerical experiments}. Clearly,  more dedicated efforts are needed to quantify the impact of these waves on the mean flow. 

In this paper, we explore the impact of vertically distributed sources of gravity waves on the middle and upper atmosphere of Earth. For this, we use the Coupled Middle Atmosphere Thermosphere-2 General Circulation Model \cite<CMAT2-GCM,>[]{Yigit.etal2009_ModelingEffects}, incorporating the whole atmosphere nonlinear gravity wave parameterization of \citeA{Yigit.etal2008_ParameterizationEffects}. We utilize the new capability of the gravity wave scheme in accounting for the effects of gravity waves associated with extra-tropospheric sources  \cite{Medvedev.etal2023_DynamicalImportance}. We explore the upper limit of the importance of secondary waves by assuming that the extra-tropospheric sources generate waves in sync with the incident spectrum. In other words, we assume that they always amplify the incident harmonics. 

The structure of the paper is as follows: The next section describes the numerical modeling approach, including Coupled Middle Atmosphere Thermosphere-2 (CMAT2) general circulation model, the whole atmosphere gravity wave parameterization, the implementation of the higher order gravity waves, and data analysis. Results are presented systematically in terms of column model simulations (Section \ref{sec:column-model-analys}), GCM simulations (Section \ref{sec:gcm-results}), and comparison of CMAT2 simulations with ICON horizontal wind measurements (Section \ref{sec:cmat2-icon}). Summary and conclusions are given in Section \ref{sec:summary}.

\section{Data \& Methods}
\label{sec:data}

\subsection{Coupled Middle Atmosphere Thermosphere-2 GCM}
\label{sec:gcm}

The Coupled Middle Atmosphere Thermosphere-2 (CMAT2) General Circulation Model (GCM), originally developed at University College London, is a first-principle hydrodynamical three-dimensional time-dependent finite difference GCM. It extends from the tropopause (100 mb, $\sim$15 km) to the upper thermosphere (300--500 km), depending on solar and geomagnetic conditions. The version used in this study includes two updates to the previous model versions \cite{Yigit.etal2009_ModelingEffects,Yigit.etal2021_EffectsLatitudeDependent}: (1) Increased horizontal resolution with longitude-latitude grid spacing of $5^\circ \times 2.5^\circ$; (2) Inclusion of the vertically distributed wave sources in the gravity wave parameterization \cite{Medvedev.etal2023_DynamicalImportance}.  The latter update will be discussed in detail in Section \ref{sec:gw-parameterization}. The other general characteristics of the GCM are the following: At the lower boundary, the model is forced by the National Centers for Environmental Prediction (NCEP) reanalysis daily mean geopotential data, filtered for planetary-scale wave numbers one to three, and the  Global Scale Wave Model(GSWM)  data, representing solar tidal forcing \cite{Hagan.Forbes2002_MigratingNonmigrating}.  In the vertical, the model has 63 pressure levels with one-third scale height resolution, except at the top 3 levels, where one-scale height resolution is used.

For upper atmospheric processes, CMAT2 incorporates several components. An empirical ionosphere model spanning the low- to high-latitudes is used in all simulations \cite{Daniell.etal1995_ParameterizedIonospheric}. Realistic magnetic field distribution is specified via the International Geomagnetic Reference Field model \cite<IGRF,>[]{Thebault.etal2015_InternationalGeomagnetic}. Thermospheric heating, photodissociation, and photoionization are calculated for the absorption of solar X-rays, extreme ultraviolet (EUV), and UV radiation between 1.8 and 184 nm using the SOLAR2000 empirical model of \citeA{Tobiska2000_StatusSOLAR2000}. Further details of these models can be found in the work by \citeA{Yigit.etal2009_ModelingEffects}.

CMAT2 has been extensively used to study vertical coupling between the lower and upper atmosphere via subgrid-scale gravity waves and solar tides, and has been validated with respect to observations and empirical models \cite{Yigit.Medvedev2009_HeatingCooling, Yigit.Medvedev2010_InternalGravity,
Yigit.Medvedev2017_InfluenceParameterized, 
Yigit.etal2009_ModelingEffects, 
Yigit.etal2012_DynamicalEffects, Yigit.etal2021_EffectsLatitudeDependent}. These studies demonstrated the suitability of CMAT2's dynamical core for modeling the dynamical and radiative processes in the middle and upper atmosphere. 

\subsection{Whole Atmosphere Gravity Wave Parameterization}
\label{sec:gw-parameterization}
The  parameterization presented in detail in the work by \citeA{Yigit.etal2008_ParameterizationEffects} is used to represent the effects of subgrid-scale gravity waves. The vertically distributed wave sources are accounted for according to the work of \citeA{Medvedev.etal2023_DynamicalImportance}. We first describe the incident wave spectrum in the lower atmosphere and then explain how the sources at other altitudes were incorporated into the model. 

\subsubsection{Incident Gravity Wave Spectrum}
\label{sec:initial-gw-spectrum}

\begin{figure}[h]
  \begin{centering}
    \includegraphics[width=1.0\textwidth,
    keepaspectratio, height=0.9\textheight]{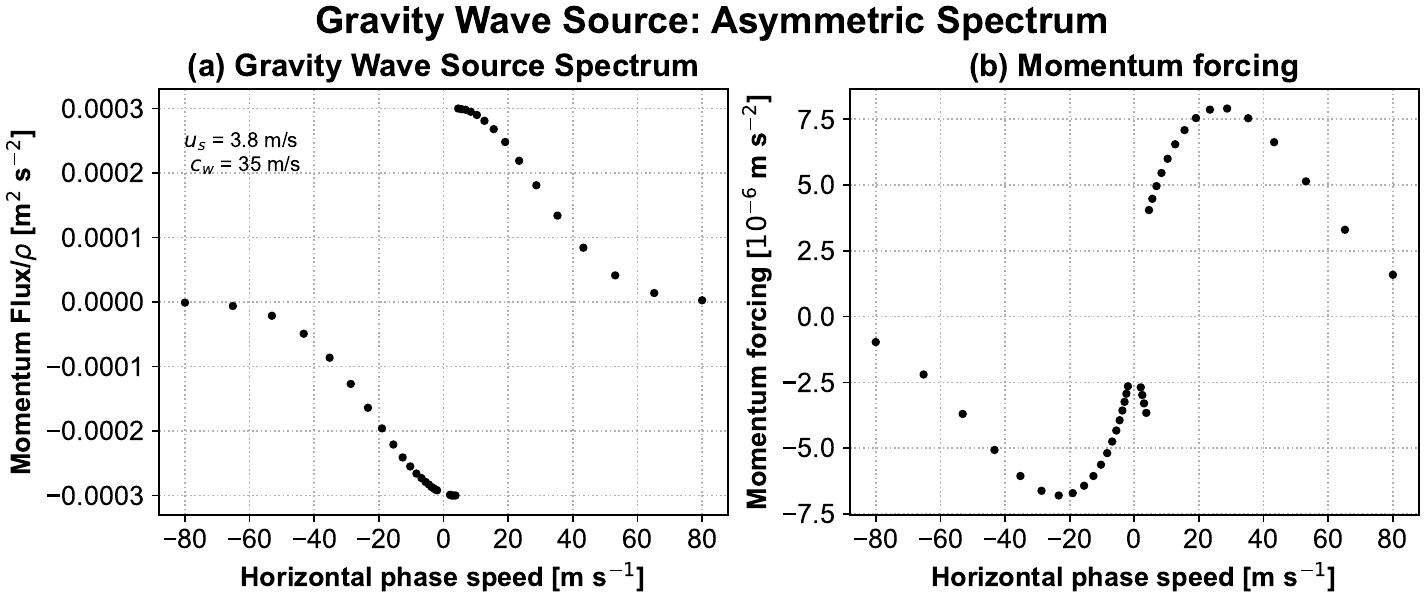}
    \caption{(a) Gravity wave spectrum at the source level  represented by a Gaussian distribution of horizontal momentum fluxes as a function of horizontal phase speeds on 2 June 2020, 0000 UT at 50$^\circ$N, 0$^\circ$ longitude (Greenwich Meridian). The source spectrum is asymmetric since it takes into account the local mean winds ($u_s =3.8$ m~s$^{-1}$ at this location and time). Panel (b) shows the associated momentum forcing (in m~s$^{-2}$) required to generate the spectrum in (a).} 
    \label{fig:spectrum_figure}
  \end{centering}
\end{figure}

The main adjustable parameters of the scheme are the characteristic horizontal wavelength $\lambda_h$ and  the incident source spectrum. We use a representative  $\lambda_h= 300$ km (i.e., horizontal wavenumber $k_h =  2\pi/\lambda_h = 2.094\times 10^{-5}$ m$^{-1}$). The  gravity wave activity at the lower boundary is specified  at the source level by a Gaussian distribution of vertical flux of horizontal wave momentum as a function of horizontal phase speeds:

\begin{equation}
  \label{eq:initial_spectrum}
  \overline{u^\prime w^\prime}_i (z_0)= \textrm{sgn} (c_i - \bar{u}_s) \;
  \overline{u^\prime w^\prime}_{max}
  \exp \Bigg[ \frac{-(c_i -\bar{u}_s)^2}{c_w}\Bigg],
\end{equation}
where the subscript $i$ denotes a gravity wave harmonic, $z_0\approx 15$ km is the source level,  $ \overline{u^\prime w^\prime}_{max}$ is the maximum amplitude of the Gaussian distribution, $c_w =35$ m~s$^{-1}$ is the half-width at half-maximum, and $c_i - \bar{u}_s = \hat{c}_i$ is the intrinsic horizontal phase speed at the source level with the source mean wind $\bar{u}_s$. The  gravity wave source spectrum and the associated momentum forcing that generates this spectrum (assuming uniform forcing between the surface and $z_0$) are shown in \Fig \ref{fig:spectrum_figure}a on 2 June 2020, 0000 UT at 50$^\circ$N, 0$^\circ$ longitude (Greenwich Meridian). The phase speed spectrum includes a total of 38 harmonics ranging from $\pm2 $ to $\pm 80 $ m~s$^{-1}$ propagating in opposite directions. The spectrum is asymmetric because the momentum flux distribution is shifted by the source winds as shown mathematically in  (\ref{eq:initial_spectrum}). If the source wind is zero, the momentum flux spectrum would be symmetric and perfectly centered around the zero speed harmonic.  The associated momentum forcing required to generate this spectrum is shown in \Fig \ref{fig:spectrum_figure}b and is calculated using Equation (20) in the work by \citeA{Medvedev.etal2023_DynamicalImportance}. Although the maximum gravity wave source amplitude  ($\overline{u^\prime w^\prime}_{max} = 3\times 10^{-4}$ m$^{-2}$ s$^{-2}$) is fixed in our study, the asymmetric source spectrum is spatially and temporally variable, since the source winds vary geographically and as a function of time.


\subsubsection{Vertical Evolution and Dissipation of Gravity Waves}
\label{sec:vert-evol-diss}

The vertical propagation of gravity waves in the atmosphere is governed by several physical processes that affect their amplitude and momentum flux. The vertical evolution of the horizontal momentum flux above the initial source is given by
\begin{equation}
  \label{eq:uw}
  \overline{u^\prime w^\prime}_i (z) = \overline{u^\prime w^\prime}_i (z_0) \frac{\rho_0}{\rho(z)} \tau_i(z),
\end{equation}
where $\rho_0 = \rho(z_0)$ is the background density at the reference  level ($z_0\approx 15$ km), $\tau_i(z)$ is the wave transmissivity of the $i$-th harmonic, and the altitude variation of the background density is
\begin{equation}
  \label{eq:rho}
  \rho(z) = \rho_0 \exp \bigg( -\int_{z_0}^z \frac{dz}{H_\rho(z)} \bigg)
\end{equation}
with the density scale height $H_\rho$
\begin{equation}
  \label{eq:Hrho}
  \frac{1}{H_\rho} \equiv \frac{Mg}{RT} + \frac{1}{T} \frac{dT}{dz},
\end{equation}
where $H = RT/Mg$ is the pressure scale height, $R=8.314$ J K$^{-1}$ mol$^{-1}$ is the universal gas constant, $g$ is gravitational acceleration, $T$ is temperature, and $M$ is molar mass.  For an isothermal atmosphere, the density scale height is equal to the pressure scale height. Variation of the density with height is one of the key parameters that controls wave amplitude growth.

Total gravity wave dissipation $\beta_{tot}$ is calculated as a superposition of the effects of nonlinear interactions ($\beta_{non}^i$), molecular diffusion and thermal conduction ($\beta_{mol}^i$), radiative damping ($\beta_{rad}^i$), and ion drag ($\beta_{ion}^i$) on each gravity wave harmonics:
\begin{equation}
\label{eq:total_dissipation_rate}
    \beta_{tot} = \sum_i \beta_{non}^i  + \beta_{mol}^i + \beta_{rad}^i +  \beta_{ion}^i
\end{equation}
The resulting gravity wave drag vector $\mathbf{a}$ is:
\begin{equation}
\label{eq:gravity_wave_drag}
    \mathbf{a} = (a_x,a_y) 
    = \bigg( 
    -\frac{1}{\rho} \frac{\partial (\rho \overline{u^\prime w^\prime}) }{\partial z} ,
    -\frac{1}{\rho}  \frac{\partial (\rho \overline{v^\prime w^\prime})}{\partial z}
    \bigg)
\end{equation}
and the resulting net heating/cooling rate $\epsilon$ is the sum of irreversible heating $E_i$ and the differential heating/cooling $Q_i$
\begin{equation}
    \epsilon_i = \underbrace{\frac{a_i (c_i -\bar{u})}{c_p}}_{E_i} + 
    \underbrace{\frac{H}{2\rho R} \frac{\partial}{\partial z} [\rho a_i (c_i - \bar{u})]}_{Q_i},
\end{equation}
where $c_p$ is the specific heat at constant pressure and $a_i$ gravity wave drag along the direction of $c_i$. 
The whole atmosphere gravity wave scheme is also applicable to other planetary atmospheres. Its most recent application in the Martian atmosphere can be found in the work of \citeA{Shaposhnikov.etal2022_MartianDust}.

\subsubsection{Implementation of Vertically Distributed Sources}
\label{sec:impl-high-order}

 The exact mechanism of  wave generation over all heights is not well known and quantified, and therefore cannot be directly implemented in the gravity wave parameterization. Instead, we use the approach introduced in the paper of \citeA{Medvedev.etal2023_DynamicalImportance}. It consists in expressing the unknown wave forcing at different altitudes in terms of the tropospheric forcing $G_{trop}$ (see \Fig \ref{fig:spectrum_figure}b) that formed the spectrum (\ref{eq:initial_spectrum}) at the source level $z_0$. While the magnitude of the non-tropospheric sources introduced in this way is given as multiples of $G_{trop}$ in the numerical experiments, we kept their  representative horizontal wavelength and spectral shape the same as for the tropospheric sources.

\subsection{Simulation Design and Experiments}
\label{sec:experiments}
In the results to be presented in this paper, the CMAT2 model has been run under different configurations of the  gravity wave scheme described above. First, we perform the benchmark simulation (or EXP0) with the sources  (\ref{eq:initial_spectrum}) prescribed at the lower boundary at 15 km.  Three further simulations, EXP1--EXP3, included sources of different intensities (with respect to $G_{trop}$) launched at different levels in addition to the waves of tropospheric origin.   EXP1 assumes the localized gravity wave sources at 90 km that are ten times stronger than those in the troposphere. EXP2 is the same as EXP1 but the localized sources are located at 50 km. These localized sources in the middle atmosphere are much stronger than the incident source in the lower atmosphere because we were exploring the upper limit of the importance of non-tropospheric sources.  In EXP3, secondary sources equal to $G_{trop}$ were placed at all levels above $z_0$. This is the case with the strongest extra-tropospheric forcing. Note that the above cases were tested by \citeA{Medvedev.etal2023_DynamicalImportance} in a column model at a representative latitude and longitude.  Here we have studied them in a fully interactive three-dimensional global model. The described runs  are listed in Table \ref{tab:model-exp}. 

To generate initial conditions for our model experiments, the CMAT2 GCM was first run in perpetual mode for 40 days for March 20, 2020, and reached steady state (EXPS). From there, the model was run in day-stepping mode  from March 21, 2020 to July, 5 2020.  Our analysis focuses   on the 30-day period (6 June--5 July 2020) around the Northern Hemisphere summer solstice, for which  data were output every three hours. The mean fields to be presented are evaluated as time and longitude averages for the selected period. For the column model tests, we  run the GCM for one time step (on 2 June 2020, 0000 UT) and illustrate gravity wave propagation and dissipation at a representative latitude and longitude. 

\renewcommand{\arraystretch}{1.3}
%
%
\begin{table}[H]
  \centering
  \caption{Description of the CMAT2 model experiments, EXPS and EXP0--EXP3. EXP0 is the benchmark run, including primary waves of tropospheric origin only. EXP1--3 are the runs with extra-tropospheric sources, i.e., with gravity wave sources localized and distributed at different heights. The benchmark case (EXP0) is the standard case with gravity waves launched at $\sim 15 $ km.}
  \begin{tabular}{m{2.0cm}||m{4.3cm}|m{3cm}|c|c}
    \toprule\centering
    \textbf{Model} \break \textbf{Experiment}
    & \textbf{Description}
    & \textbf{Source location}
    & $G_{trop}$ 
    & \textbf{Simulation Dates} \\
    \midrule\midrule
    EXPS &   Spinup run  & No gravity waves  & -- & 20 March 2020  \\
    \hline
    EXP0 &   Benchmark run \break  (tropospheric sources only) & 15 km & -- 
    &  \multirow{4}{*}[-2.em]{21 March -- 5 July 2020}   \\
    \cline{1-4}
    EXP1 &   Tropospheric \break + localized sources           &  15 km and 90 km & $10\times $\\
    \cline{1-4}
    EXP2 &   Tropospheric \break + localized sources           &  15 km and 50 km & $10\times $   \\
        \cline{1-4}
    EXP3 &   Tropospheric \break + uniform sources above       &  15 km upward    & $1\times $  \\
    \bottomrule
  \end{tabular}
  \label{tab:model-exp}
\end{table}

\subsection{Method of ICON Data Analysis and Comparison with CMAT2 GCM}

We use  zonal and meridional winds measured by the satellite to compare with our simulations. For this, a validation framework was developed to evaluate the performance of the CMAT2 model in simulating thermospheric winds through direct comparison with NASA's ICON (Ionospheric Connection Explorer) satellite observations from the MIGHTI instrument \cite{Immel.etal2018_IonosphericConnection, Englert.etal2017_MichelsonInterferometer}. The framework integrates satellite data processing, spatial-temporal matching algorithms, and statistical analysis to quantify model performance across different wind components and altitude regions. ICON was launched on 10 October 10 2019 and operated till 25 November 2022,  observing the low-latitude thermosphere-ionosphere system above 90 km in unprecedented detail, concentrating on the understanding the complex interactions between Earth's upper atmosphere and the geospace environment \cite{Immel.etal2018_IonosphericConnection}.

The MIGHTI (Michelson Interferometer for Global High-resolution Thermospheric Imaging) instrument aboard NASA's ICON provides  measurements of thermospheric neutral winds through observations of naturally occurring atomic oxygen airglow emissions. We use the Level 5 Neutral Wind data product, which represents altitude profiles of the full horizontal wind vector (both zonal $u$ and meridional $v$ components) in geographic and geomagnetic coordinate systems. This product is derived by combining line-of-sight wind measurements from the MIGHTI-A and MIGHTI-B sensors, which view the atmosphere at approximately perpendicular angles ($\sim$45$^\circ$ and $\sim$135$^\circ$ relative to the spacecraft velocity vector).

MIGHTI observes two distinct emissions: the green line (557.7 nm) and the red line (630.0 nm). The green line observations cover the altitude range of 90--170 km during daytime and 90--105 km at night, while the red line provides measurements from 170--300 km during daytime and 210--300 km at night. The data are binned to a resolution of 5 km for green line and 30 km for red line measurements to improve statistical quality while meeting ICON science requirements \cite{Harding.etal2017_MIGHTIWind}. The satellite data processing component filters measurements based on wind quality flags and excludes measurements with absolute wind speeds exceeding 350 m~s$^{-1}$.

The core validation mechanism matches satellite measurements with corresponding model output points. For spatial matching, each satellite measurement is associated with the nearest model grid point, with measurements that are more than  half the grid distance from any grid point excluded from the analysis. Temporal matching groups satellite measurements into time bins, which are then paired with the closest model time step within a configurable threshold (default is 3 hours). This approach accommodates differences in temporal sampling between the model and observations. To account for different vertical resolutions in the ICON measurements, the framework implements height binning: 5 km bins for greenline and 30 km bins for redline measurements. Additionally, the system incorporates solar zenith angle filtering capabilities, allowing separate analysis of daytime, nighttime, and terminator conditions.

ICON/MIGHTI winds have been extensively used to study coupling and interaction processes in the atmosphere, specifically to study semidiurnal-induced vertical coupling \cite{Forbes.etal2022_VerticalCoupling}, global-scale wave coupling \cite{Gasperini.etal2023_IonospherethermosphereCoupling}, equatorial ionospheric convective instability \cite{Hysell.etal2023_ForecastingEquatorial}, impact of volcanic eruption on the ionospheric dynamo \cite{Harding.etal2022_ImpactsJanuary}, the response of the thermospheric circulation to a major SSW \cite{Yigit.etal2024_ObservationVertical} and to Hurricane Grace \cite{Gann.Yigit2024_IonosphericThermospheric}.

\section{Column Model Analysis and Prelude to Gravity Wave Propagation}
\label{sec:column-model-analys}

Numerous insights into gravity wave propagation from the lower atmosphere to the upper atmosphere can be gained by using a column model version of the whole atmosphere gravity wave parameterization. For this, we analyze gravity wave propagation and dissipation on 2 June 2020 at 0000 UT at a representative latitude ($\theta = 75^\circ$N) and longitude ($\phi = 0^\circ$, Greenwich Meridian) for a single probe harmonic propagating in the eastward direction ($c_i=80$ m s$^{-1}$) in the presence of a broad spectrum of gravity waves described by   (\ref{eq:initial_spectrum}). Results are generated for diagnostic purposes by running the CMAT2-GCM for one time step and outputting the background atmospheric fields and gravity wave parameters. We use by default the asymmetric source spectrum (\ref{eq:initial_spectrum}), in which the gravity waves are launched in the direction of the source horizontal winds.

\begin{figure}[h!]
  \includegraphics[width=1\textwidth,
  keepaspectratio, height=0.9\textheight]{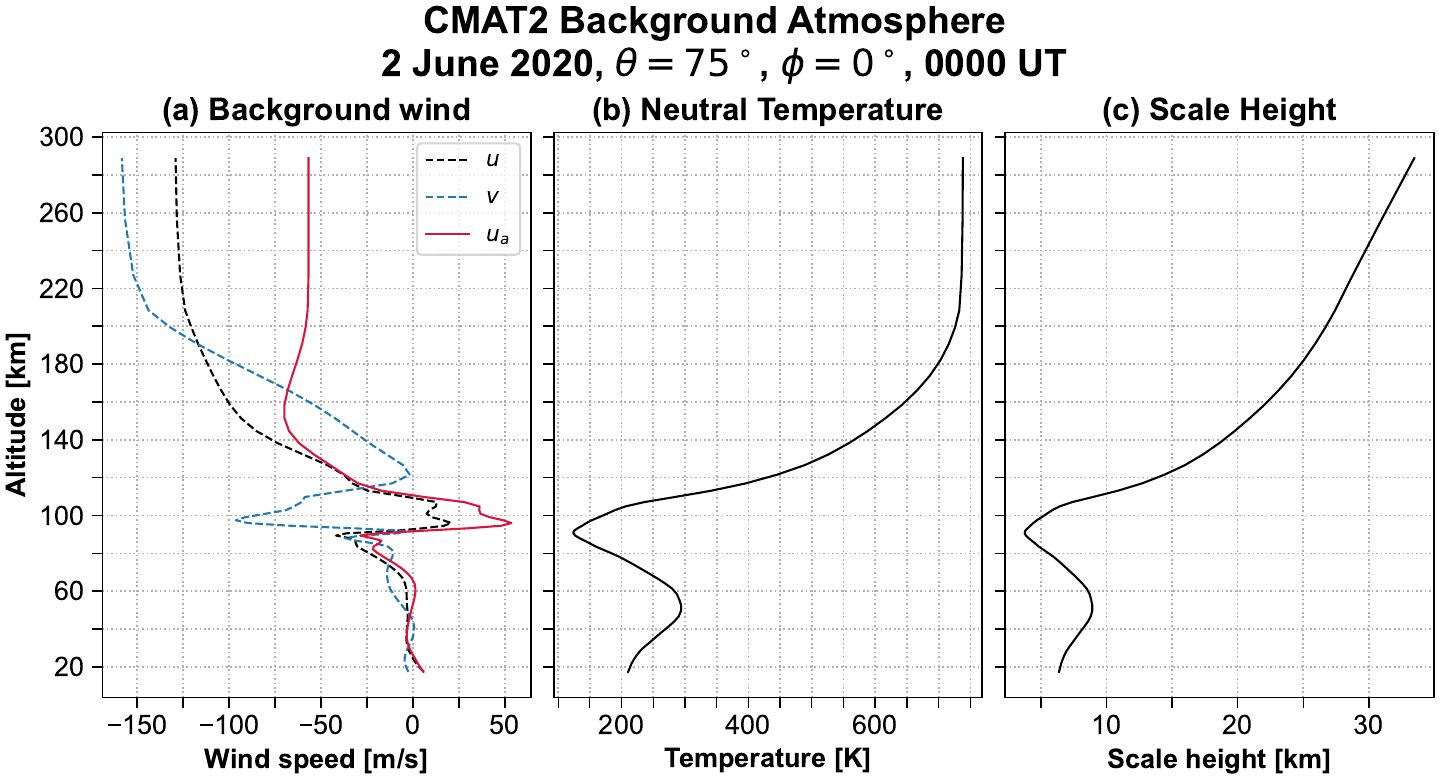}
  \caption{The background atmosphere as simulated by the CMAT2 GCM at a representative location and time: Latitude $\theta =75^\circ$N, longitude $\phi = 0^\circ$ (Greenwich Meridian), 2 June 2020, 0000 UT. (a) Background horizontal wind components, (b) neutral temperature $T$, (c) pressure scale height $H$. In (a), $u $ and $v$ denote the zonal and meridional components of the horizontal wind velocity $\mathbf{u}$, and $u_a $ is the  projection of the upper level horizontal winds on the direction of wave propagation, which coincides with the wind direction at the launch height.}
  \label{fig:background-wind}
\end{figure}

The vertical profiles of the background winds, neutral temperature and pressure scale height are seen in \Fig \ref{fig:background-wind} on 2 June 0000 UT in the chosen location. For the representative background conditions, the zonal wind ($u$) alternates its direction with altitude. It is westward in the middle atmosphere, peaking with $\sim$--45 m s$^{-1}$ around 95 km, reverses its direction around 100 km to eastward increasing up to 25 m s$^{-1}$ and reverses back to westward direction above 110 km, growing in magnitude and reaching --125 m s$^{-1}$ above 200 km. The meridional wind ($v$) is overall southward and has a stronger magnitude than the eastward wind above 95 km. The projection on the direction of wave propagation wind, i.e., the anisotropic wind ($u_a$), maximizes around 100 km with 50 m s$^{-1}$ and reverses its direction above 120 km varying from 50 to --75 m s$^{-1}$ in the upper atmosphere. The temperature variations demonstrate the characteristic profile of the neutral temperature with the alternating temperature gradients between the different "pauses", becoming isothermal in the upper thermosphere, reaching $\sim$ 750 K, which is typical for solar minimum conditions. The scale height closely follows the temperature profile as expected from the linear relation between scale height and temperature. 
  
\begin{figure}[h!]
  \centering
  \hspace*{-1.3cm}
  \includegraphics[width=1.2\textwidth]{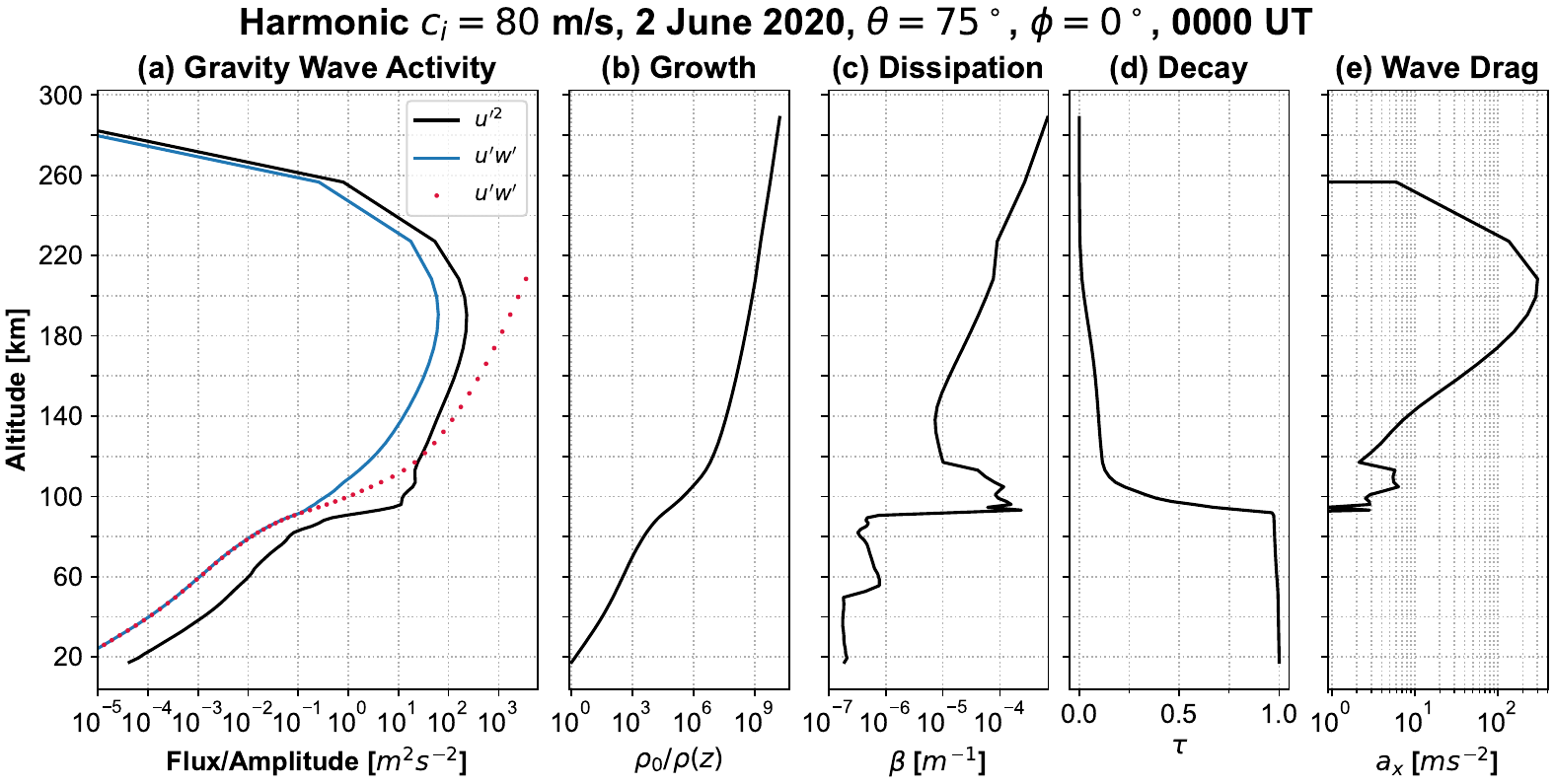}
  \caption{Instantaneous column model results for  the incident test harmonic $c_i = 80$ m~s$^{-1}$ on 2 June 2020 0000 UT, taking into account nonlinear interactions with a broad spectrum of waves, at a representative grid point  ($\theta = 75^\circ$, $\phi = 0^\circ$): (a) gravity wave variance $\overline{u^{\prime 2}}$ (black) and horizontal momentum flux $\overline{u^\prime w^\prime}$ (blue) in m$^{-2}$~s$^{-2}$; the  wave flux per unit mass in case of conservative propagation (red dotted) is added. (b) Background density variation relative to the source level density $\rho_0$, which is representative of the wave growth; (c) total gravity wave dissipation rate $\beta$; (d) gravity wave transmissivity $\tau$; (e) zonal gravity wave drag $a_x$ in m~s$^{-1}$~day$^{-1}$. The incident gravity wave spectrum is the same as in Figure \ref{fig:spectrum_figure} with the maximum source flux of $\overline{u^\prime w^\prime}_{max} = 3 \times 10^{-4}$~m$^{-2}$~s$^{-2}$ and the characteristic horizontal wavelength $\lambda_h = 300$ km. The source  flux associated with the test harmonic is $\overline{u^\prime w^\prime}_i (z_0) \approx 0.329 \times 10^{-5}$ m$^2$~s$^{-2}$. Extra-tropospheric sources are not included.} 
    \label{fig:column_model-c-80}
  \end{figure}

Next we study in this background atmosphere the growth, decay/dissipation of the probe harmonic ($c_i = 80$ m s$^{-1}$), and the resulting wave activity and effects in \Fig \ref{fig:column_model-c-80}. The gravity wave activity in terms of the horizontal momentum flux $u^\prime w^\prime_i$ and variance $u^{\prime2}_i$ (\Fig \ref{fig:column_model-c-80}a)  grows nearly exponentially by several orders of magnitude from its source to about 100 km, as revealed by the comparison to the idealized case of dissipationless (i.e., conservative) wave growth (red dotted). From 100--200 km the growth slows down, and above 200 km the wave activity decreases strongly.   

The upward propagation of a gravity harmonic is determined primarily by two factors: (1) amplitude growth and (2) wave dissipation and/or critical level filtering. How much the amplitude or flux/variance of a wave will grow depends on the change of the background density with altitude with respect to the reference density, i.e., $\rho_0/\rho(z)$. While the growth rate due to the density decrease is the same for all gravity wave harmonics, the dissipation rates $\beta^i$ (and critical level filtering) of the individual harmonics differ from each other. This is why different harmonics propagate to different altitudes with varying amplitudes/fluxes. For our case, the background density scaled by the source density increases from unity at the source level to $\sim 10^{10}$ at 260 km (\Fig \ref{fig:column_model-c-80}b), which suggests that in the absence of dissipation and critical level filtering the horizontal momentum flux would increase by 10 orders of magnitude from the tropopause to the upper thermosphere in order to conserve wave energy. The growth factor follows from (\ref{eq:uw}) and (\ref{eq:rho})
\begin{equation}
  \label{eq:density-growth}
\frac{\rho_0}{\rho(z)}
  \sim \exp \bigg(\int \frac{dz}{H_\rho} \bigg), 
\end{equation}
which suggests that the density does not grow at a fixed 
exponential rate (as seen in panel b), because the density scale height $H_\rho$ depends on both the pressure scale height and the vertical temperature gradient. Lighter species prevail in the thermosphere compared to the middle atmosphere, thus the molecular mass (or molar mass) is smaller there than at lower altitudes. Together with a greater thermospheric temperature gradient, this leads to a weaker wave growth above 130 km compared to the middle atmosphere. 

A gravity wave amplitude  cannot grow indefinitely. In the middle atmosphere and thermosphere, dissipative processes take effect, provided that a given harmonic has survived critical level filtering at lower altitudes. The total dissipation rate (\ref{eq:total_dissipation_rate})  resulting from nonlinear damping, molecular viscosity, and ion drag, etc., is seen in \Fig \ref{fig:column_model-c-80}c. Overall, the dissipation rate grows exponentially as well, increasing by  four orders of magnitude from the troposphere to the upper thermosphere.  The wave dissipation in the middle atmosphere and the local dissipation maximum   around 100 km are due to nonlinear interactions ($\beta_{non}^i$) with other gravity wave harmonics.  The thermospheric dissipation increasing by two orders of magnitude above 120 km is primarily due to the molecular viscosity and thermal conduction $\beta_{mol}^i$ and ion-neutral frictional damping $\beta_{ion}^i$, where the latter is playing a secondary role. The effect of dissipation is incorporated in the wave transmissivity $\tau_i$ (\Fig \ref{fig:column_model-c-80}d), which is unity up to about 60 km, meaning that the probe wave grows in amplitude/flux nearly conservatively. Above 60 km, nonlinear damping gradually takes  effect, since other harmonics in the spectrum grow in amplitude as well and start impinging on the given test harmonic through nonlinear interactions. Around 100 km, more than a factor of two decrease occurs in $\tau_i $ owing to the combined effects of the local maximum in nonlinear damping and decrease in the intrinsic phase speed $\hat{c}_i = c_i - u_a$ of the test harmonic. In the thermosphere above 180 km, $\tau_i $ goes to zero as a consequence of the viscous dissipation $\beta_{mol}^i$ leading to the absorption of the wave by the background atmosphere.

Overall, the vertical evolution of the vertical flux of  horizontal wave momentum    $\overline{u^\prime w^\prime}_i (z)$ (i.e., gravity wave activity) is controlled by the competition between the growth due to the decreasing background density and the damping due to dissipative processes. When the growth rate exceeds  dissipation, the wave activity grows until reaching the lower thermosphere. Above this level,  wave growth slows down, and above 200 km, wave activity decreases exponentially at a rate increasing with altitude. Increased  momentum deposition occurs at altitudes where the wave carries a sufficient momentum flux in the presence of dissipation. Around 100 km, there is a local maximum of gravity drag of the order of 10 m~s$^{-1}$~day$^{-1}$, but the the global maximum of drag occurs at 200 km with up to 300 m~s$^{-1}$~day$^{-1}$. Although the dissipation is much stronger above 200 km, the wave has transferred much of its momentum to the background atmosphere below, thus the drag decreases rapidly above 200 km.


\section{Global Modeling of  Gravity Wave Effects}
\label{sec:gcm-results}
\subsection{Monthly Mean Background Fields and Gravity Wave Effects}
Preliminary tests in a column model have shown that  gravity wave sources  at higher altitudes contribute negligibly to wave effects compared to those  generated in the troposphere  \cite{Medvedev.etal2023_DynamicalImportance}. However, such steady-state results with a column model  provide only a limited view of gravity wave effects in a realistic atmosphere. Since wave propagation and dissipation are highly variable, three-dimensional time-dependent simulations  are needed to gain deeper insight into the relative dynamical role of various sources. For this, we performed GCM simulations as described in section~\ref{sec:experiments}. 

\begin{figure}[t!]
    \centering
    \vspace{-1.4cm}
    \hspace*{-1cm}
    \includegraphics[width=1.1\linewidth]{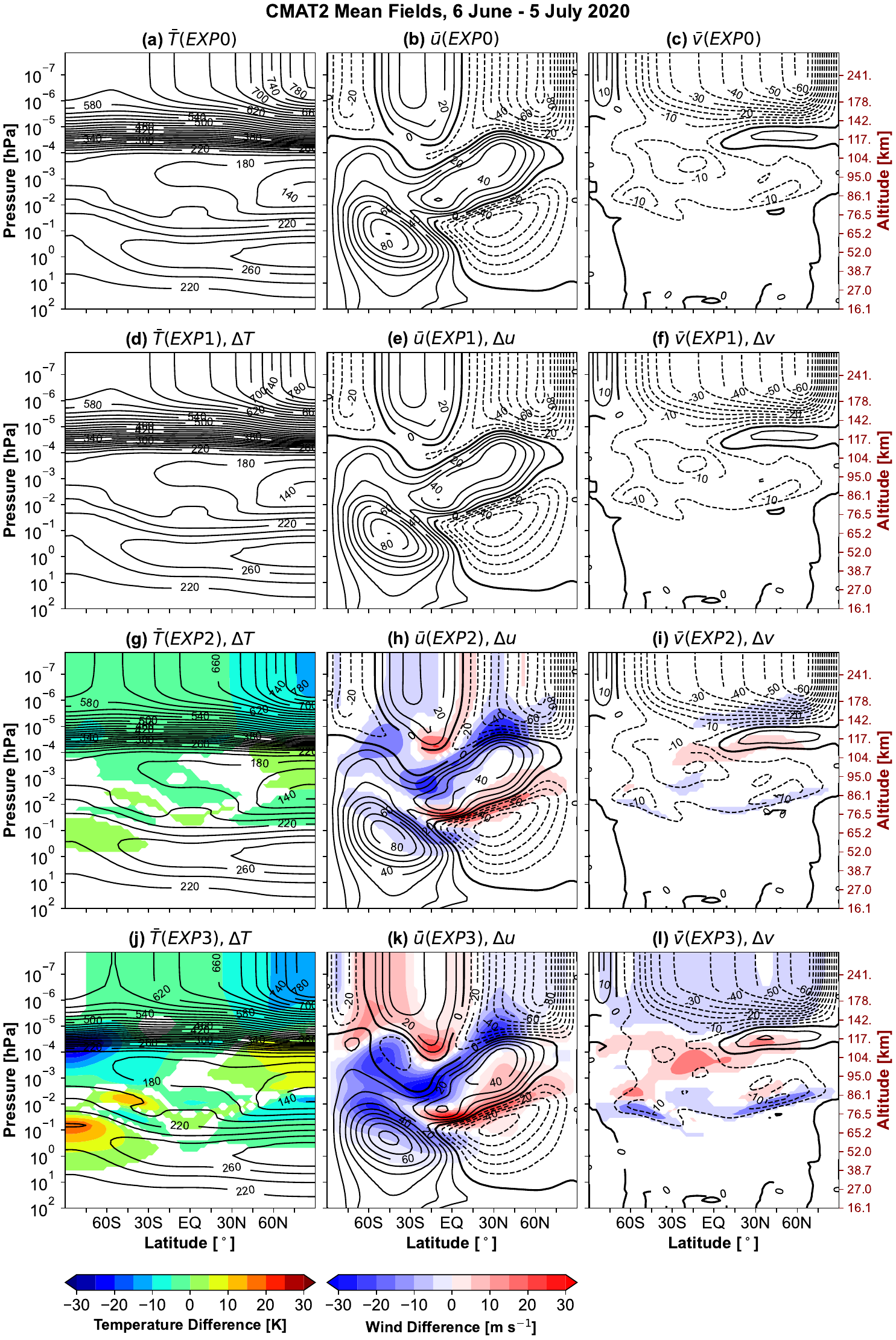}
    \caption{Pressure-latitude distributions of the  fields simulated with CMAT2 for the Northern Hemisphere summer solstice conditions: (a) Neutral temperature $T$, (b) zonal wind $u$, (c) meridional wind $v$. The fields are averaged longitudinally and temporally between June 6 and July 5. Shown are the simulations including only the tropospheric sources (EXP0), with added localized sources at 90 km (EXP1) and 50 km (EXP2), and with sources distributed over all heights (EXP3). Solid and dashed lines for the zonal and meridional winds represent eastward/northward and westward/southward directions, respectively. The differences between the runs EXP1--EXP3 and the benchmark EXP0 are shown in color shading. The differences for the zonal and meridional use the same color scale. The mean geopotential height is shown on the right.}
    \label{fig:mean_fields_all_pres}
\end{figure}

The time- and zonal averaged temperature $\bar{T}$, zonal wind $\bar{u}$, and meridional wind $\bar{v}$ are plotted in columns (from left to right) in \Fig \ref{fig:mean_fields_all_pres}. The four simulations  (EXP0--EXP3) are shown in rows with the benchmark run EXP0 at the top row. The differences between  the simulations with the non-tropospheric sources and EXP0  are shown in color shading (Panels d-l)  for the mean temperature $\Delta \bar{T}$, zonal wind $\Delta \bar{u}$ and meridional wind $\Delta \bar{v}$. 

Overall, the simulated background mean fields are consistent with conditions during the Northern hemisphere summer solstice.  The reversed mesopause temperature gradient is reproduced, with a mesopause temperature of $\sim$140 K situated around 90 km near the summer pole. This is a dynamically driven phenomenon \cite{Holton1983_InfluenceGravity}. The peak winter westerly and summer easterly stratospheric jets of 90 m~s$^{-1}$ and --50 m~s$^{-1}$, respectively, and the reversal of the mean zonal wind with more than 40 m s$^{-1}$ above 85--90 km are in a good agreement with the observed wind climatologies \cite{Swinbank.Ortland2003_CompilationWind}. The reversal of the mean meridional flow in the lower thermosphere around 110--120 km from a summer-to-winter to a winter-to-summer flow exceeding 5 m~s$^{-1}$ is also reproduced. In the thermosphere above 130 km, the meridional flow  returns back to the summer-to-winter flow, reaching --60 m s$^{-1}$ and decreasing equatorward. It is  maintained primarily by the pressure gradient force produced by the differential heating between summer and winter. 


\Fig \ref{fig:mean_fields_all_pres} (panels d--l) show the results of simulations EXP1 to EXP3 with wave sources introduced above the reference height $z_0$. The localized source for EXP1 was placed in the mesopause region, an area known for strong gravity wave breaking/dissipation. The source itself was set ten times stronger than primary tropopause sources (i.e., $10\times G_{trop}$). Yet EXP1 shows virtually no difference from EXP0, which has only primary tropospheric sources. This suggests that secondary waves originating in the mesopause region are unlikely to significantly impact the zonally averaged thermospheric circulation. Waves launched at 50 km (EXP2) have five more scale heights to grow than those launched at 90 km, and with ten times the amplitude of tropospheric sources, some changes are seen (shown in color shading). For the mean zonal winds, the changes are generally found around the regions of wind reversals over a wide range of altitudes in the upper mesosphere and thermosphere. 

The largest changes (of about three times larger than in EXP2) are produced in EXP3 (panels j--l), in which sources equal in strength to those in the entire troposphere were placed at every level in the middle atmosphere. Note that there is currently no observational evidence to support such strong sources in the middle atmosphere, therefore the results should be viewed only as an upper limit estimate of possibly missing sources. Primarily, the changes are seen in the mean zonal winds, where the differences are up to $\pm 30$ m~s$^{-1}$. They show a clear hemispheric asymmetry. In the Southern Hemisphere, the winter westerlies are slower (shown by the westward wind change) between 60--130 km (Panels h,k). In the Northern Hemisphere, the mean zonal wind reversal is enhanced. Around the mesospheric mean zonal wind reversal ($\sim 85$--$90$ km at midlatitudes), the summer easterly stratospheric jet below is slower, while eastward wind reversal above is stronger. Around the lower thermospheric wind reversal  ($\sim 120$ km at midlatitudes), the thermospheric circulation is more westward above, while the eastward wind reversal is slower below. This effect shifts the upper mesospheric eastward wind reversal to lower altitudes to some degree. The addition of waves from non-tropospheric sources clearly enhances the mean meridional circulation globally. The main effects are seen in the enhancement of the winter-to-summer reversed circulation around 120 km in the Northern high-latitudes and of the summer-to winter flow in the upper mesosphere at both midlatitudes.

\begin{figure}[t!]
\vspace{-1.4cm}
    \centering
    \hspace*{-1cm}
    \includegraphics[width=1.0\textwidth]{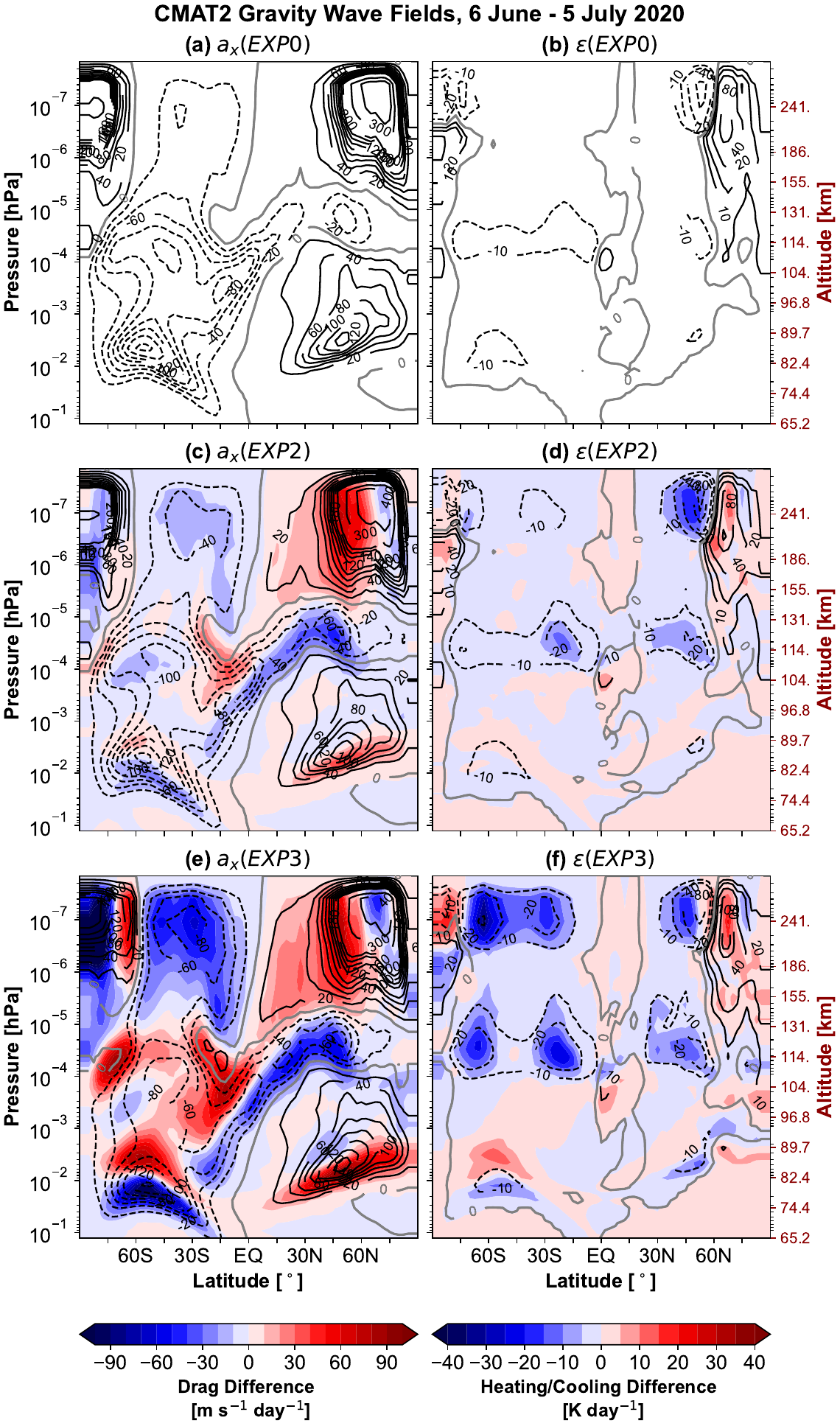}
    \caption{Same as in Figure \ref{fig:mean_fields_all_pres} but for the (a) zonal gravity wave drag $a_x$ and (b) total gravity wave heating/cooling $\epsilon$. The contour intervals for the drag are in 20 m s$^{-1}$ day$^{-1}$ within $200$ m s$^{-1}$ day$^{-1}$. Additionally 300 and 400 m s$^{-1}$ day$^{-1}$ levels are drawn. The heating/cooling contour levels are $0, \pm 10, \pm 20, \pm 40, \pm 80, \pm 100$ K day$^{-1}$. The shading intervals are 10 m s$^{-1}$ day$^{-1}$ for the gravity wave drag difference and 5 K day$^{-1}$  for the net heating/cooling difference.}
    \label{fig:gw_fields_all_pres}
\end{figure}

The primary dynamical force controlling mesospheric and lower thermospheric circulation is the deposition of gravity wave momentum, which results from the divergence of the vertical flux of  horizontal wave momentum. 
\Fig \ref{fig:gw_fields_all_pres} presents the distribution of the mean zonal gravity wave drag $a_x$ as well as the total mean gravity wave heating and cooling rates $\epsilon$ in the same manner as the mean background fields (Figure \ref{fig:mean_fields_all_pres}). We left out the results for the localized sources situated at 90 km (EXP2), because no discernible difference was seen compared to the benchmark simulation EXP0. The  drag  is concentrated in three altitude regions:  the upper mesosphere, lower thermosphere, and upper thermosphere, where it maximizes in magnitude globally. Up to the lower thermosphere peak, gravity wave dynamical effects are found at midlatitudes. In the upper thermosphere, the eastward drag  maximizes at mid-to-high-latitudes with 400 m s$^{-1}$ day$^{-1}$ and 300 m s$^{-1}$ day$^{-1}$ poleward of $45^\circ$N and $60^\circ$S, respectively. In the upper mesosphere, it is  opposite  to the mean flow in both hemispheres, i.e., it is westward in the winter and eastward in the summer. It peaks at midlatitudes with a magnitude of more than 120 m s$^{-1}$ day$^{-1}$. In the Northern midlatitude lower thermosphere around 120 km ($10^{-5}$ hPa), a region of westward drag of up to 40--60 m s$^{-1}$ day$^{-1}$ is found. In the Southern winter hemisphere, the drag is westward throughout the atmosphere, except at high-latitudes poleward of $60^\circ$S. In the upper thermosphere, gravity wave drag exceeds a few hundred m s$^{-1}$ day$^{-1}$ in both hemispheres, with stronger Northern summer drag spreading over a broader range of latitudes and peaking with more than 400 m s$^{-1}$ day$^{-1}$ than the Southern winter hemisphere drag. In general, the main effect of the exaggerated sources in the mesosphere and lower thermosphere is to amplify the effects of waves originating in the troposphere. This implies that the current drag schemes likely capture the main impact of gravity waves, and do not need to explicitly include secondary waves. 

Dissipating gravity waves can produce local heating and cooling \cite{Medvedev.Klaassen2003_ThermalEffects}. Overall, the regions affected by gravity wave thermal effects coincide with  regions experiencing  strong gravity wave drag. As demonstrated in previous GCM work \cite{Yigit.Medvedev2009_HeatingCooling}, gravity waves preferentially cool the atmosphere above the upper mesosphere. The cooling rates in the mesosphere maximize around 85 km and 120 km with with a rate of --10 K day$^{-1}$. In the thermosphere, thermal effects are concentrated at higher latitudes where they  cool the regions above the areas of heating. The typical heating and cooling rates are around --80 to 80 K~day$^{-1}$. 

It is virtually impossible to separate the effects of waves from tropospheric and non-tropospheric sources in observations.  Do  waves excited in the middle and upper atmosphere increase or decrease the effects of waves of tropospheric origin? We can address this question with simulations  by considering the contributions from waves above the troposphere.  The results are presented by color shading in difference plots  for drag in Figures \ref{fig:gw_fields_all_pres}c,e and for the net heating/cooling rate in Figures  \ref{fig:gw_fields_all_pres}d,f. Overall,  gravity waves from the non-tropospheric sources enhance the effects of those from the troposphere by up to 100 m s$^{-1}$ day$^{-1}$ in most regions, where   gravity wave drag is strong, especially in the mesosphere in both hemispheres, in the Northern lower thermosphere around 120 km, in the Southern polar and midlatitude thermosphere, and Northern midlatitude thermosphere. To a minor extent, they also counteract the effects of  waves coming from the troposphere, especially in the midlatitude upper mesosphere around 80--90 km and low-latitude lower thermosphere around 95--105 km in the Southern hemisphere. Overall, the additional waves clearly enhance the heating and cooling rates  in the lower and upper thermosphere by up to $\pm 40$ K day$^{-1}$. The results also clearly demonstrate that  gravity waves launched at all altitudes (EXP3) produce greater dynamical and thermal effects than those generated by localized sources  at higher altitudes (EXP2). This is partly due to the fact wave sources between 15 km and 50 km have up to 4 scale heights in which to grow.


\subsection{Longitudinal Variability}
\begin{figure}[t!]
    \centering
    \hspace*{-1.5cm}
    \includegraphics[width=1.25\textwidth]{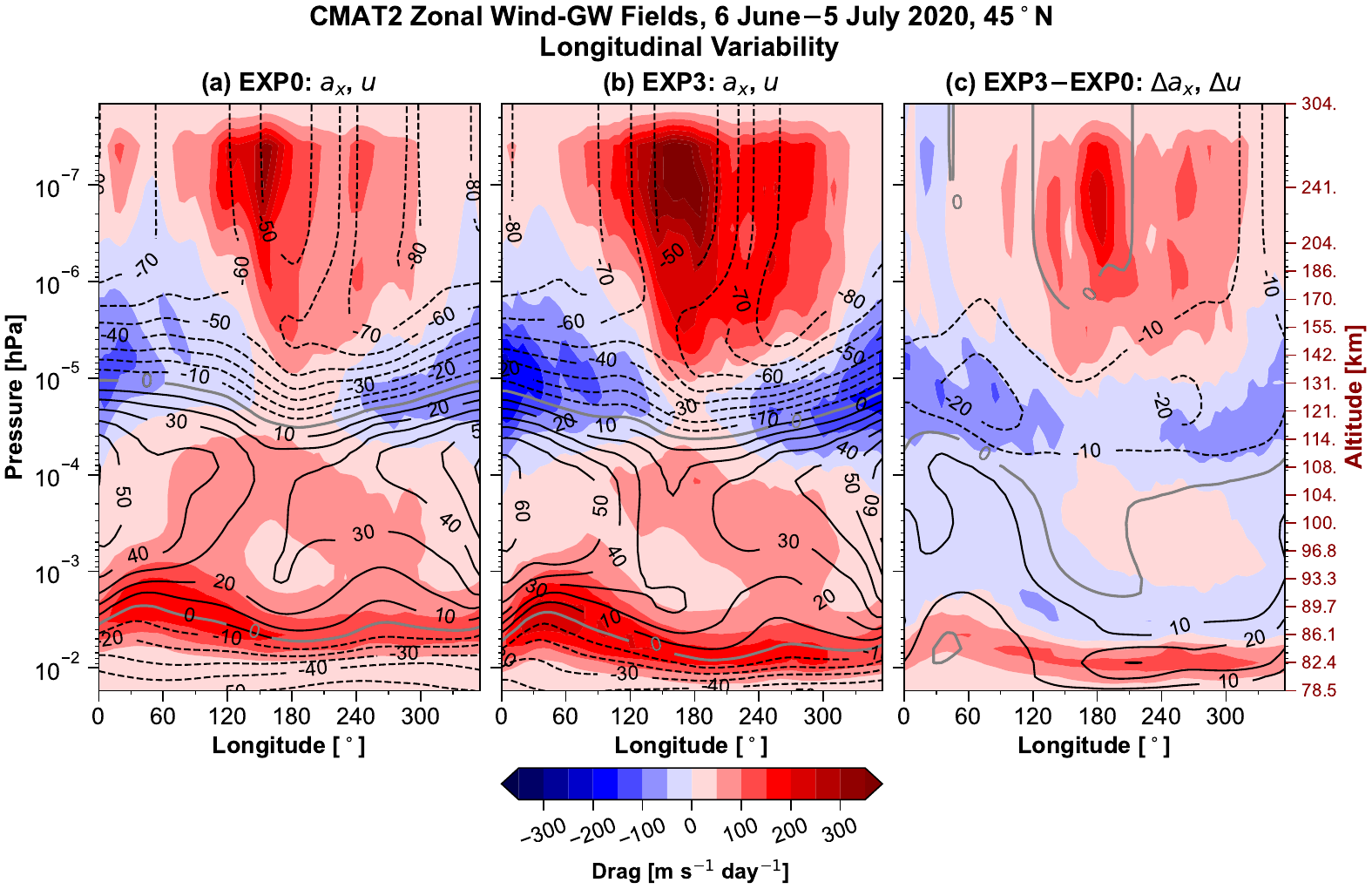}
    \caption{Longitudinal variability of the zonal wind and zonal gravity wave drag from the upper mesosphere to the thermosphere shown as pressure-longitude cross sections of the time-averaged (6 June--5July) zonal wind ($u$, contours) and zonal gravity wave drag ($a_x$, shading) at 45$^\circ$N as simulated including only tropospheric sources (EXP0, a) and including sources distributed over all heights  (EXP3, b). The difference fields (EXP3$-$EXP0) are shown in panel c. The mean geopotential height in km at chosen pressure levels are shown on the right in the red vertical axis. Drag and wind intervals are 50 m s$^{-1}$ day$^{-1}$ and 10 m s$^{-1}$, respectively. In panels a and b, red and blue shadings are for the eastward (positive) and westward (negative) gravity wave drag, respectively, and solid and dashed black contour lines are for the eastward and westwards winds, respectively. The zero wind is plotted in thick gray contour line. In panel c red and blue shadings are for eastward and westward difference relative to the benchmark simulation (EXP0). 0\degree~longitude is the Greenwich Meridian.}
    \label{fig:f6_wind_gw_pres_lon}
\end{figure}

Our analysis of monthly mean fields and the gravity wave effects has shown that non-tropospheric sources have the greatest potential impact  around the midlatitudes of both hemispheres, particularly in the run with  sources distributed at all heights (Figures \ref{fig:mean_fields_all_pres} and \ref{fig:gw_fields_all_pres}). We therefore focus on the zonal wind field and  dynamical effects of gravity waves at a representative Northern midlatitude in order to study the contribution of different sources in more detail. \Fig \ref{fig:f6_wind_gw_pres_lon} presents the pressure-longitude distribution in the upper mesosphere and thermosphere ($z >78.5$ km) of the time-averaged (from 6 June--5 July 2020) zonal wind (black contours) and zonal gravity wave drag (shading) at 45$^\circ$N  simulated in EXP0 and EXP3. The difference between them  is plotted to demonstrate the changes produced by extra-tropospheric sources of waves. Additionally, the mean geopotential height at chosen pressure levels is shown on the right. 

Since we averaged over time, local time effects are essentially removed. We concentrate on the region above the upper mesosphere, where wave magnitudes are greater. The dependence of the thermospheric gravity wave drag on longitude is conspicuous. The zonal drag maximizes at longitudes between 120\degree  and 210\degree at an altitude of around 240 km with a drag of more than 300\drag, acting against the  mean zonal wind of --50\wind in that region. This suggests that a broad spectrum of gravity waves with eastward phase speeds can propagate to these altitudes, dissipate via molecular diffusion and thermal conduction, and produce wave-induced body forcing of the flow. More eastward propagating gravity waves are able to survive critical level filtering and reach higher altitudes, because the stratospheric mean winds are easterly here and the mesospheric and lower thermospheric eastward wind reversal of up to 30\wind~ is relatively weak at around 180\degree.

The strong waves from extra-tropospheric sources imposed in EXP3 enhance the thermospheric gravity wave drag above 150 km, peaking around 240 km. They displace the reversal of the mesospheric winds to slightly lower altitudes and decelerate the eastward wind in the lower thermosphere around 180\degree~ thereby providing more favorable upward propagation conditions for gravity waves with eastward phase speeds at these longitudes. Enhanced lower thermospheric westward drag (panel c) outside the 150--180\degree~ longitude sector around 120--130 km can be explained by the deceleration of the westward upper mesospheric winds. This deceleration enables more westward gravity wave harmonics to propagate to higher altitudes, where they are more susceptible to viscous dissipation ($\beta_{mol}$) due to their decreasing intrinsic phase speeds between 120 and 155 km in the region of westward zonal winds.  

\begin{figure}[t!]
    \centering
    \hspace*{-1.5cm}
    \includegraphics[width=0.9\textwidth]{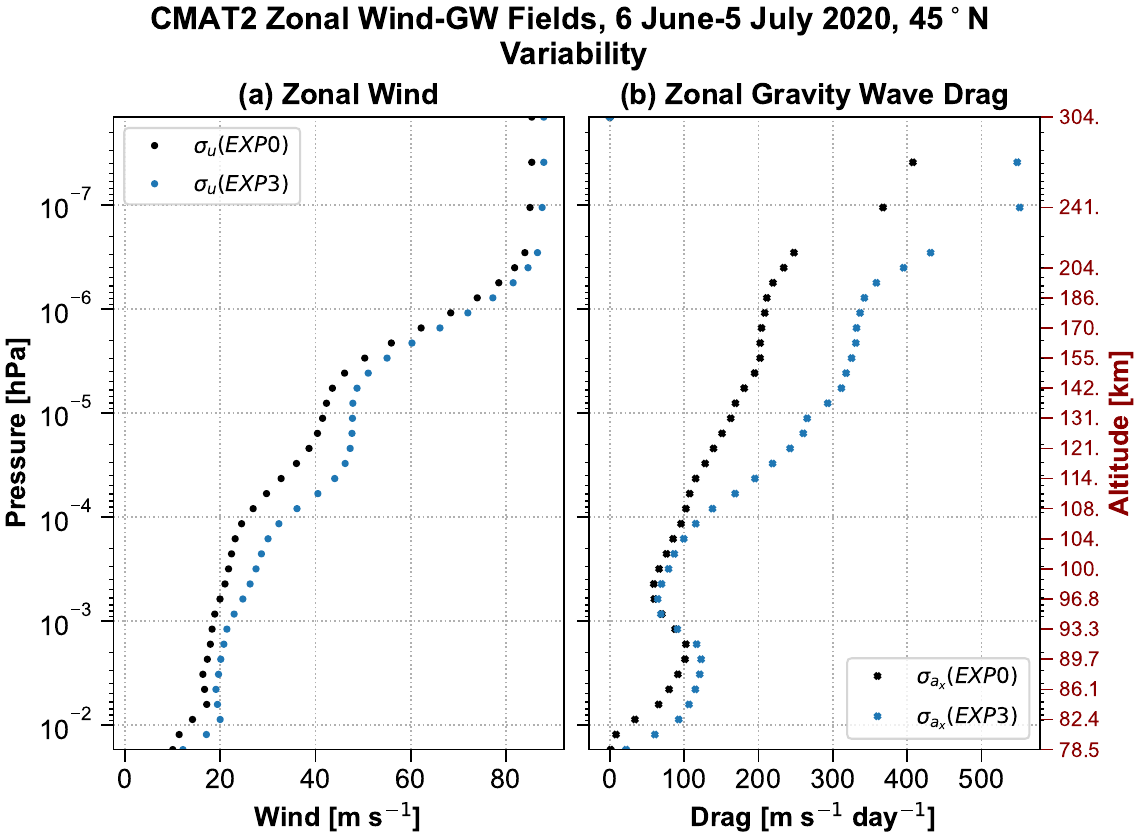}
    \caption{Standard deviation of the pressure-longitude distribution of the mean zonal wind and mean zonal gravity wave drag presented in Figure \ref{fig:f6_wind_gw_pres_lon}.}
    \label{fig:f7_wind_gw_var_mean}
\end{figure}

Gravity wave propagation into the thermosphere and the resulting dissipation are highly variable depending on the background atmosphere, wave dissipation, and critical filtering. How much do  waves of extra-tropospheric origin influence the thermospheric wind and wave variability? \Fig \ref{fig:f7_wind_gw_var_mean} studies this  in terms of the standard deviation of the pressure-longitude distributions of the fields from 6 June to 5 July at 45\degree N. This approach provides a proxy for the average longitudinal variability induced by gravity waves. In general, the longitudinal variability of both the zonal wind and drag increases with altitude. This is because gravity wave drag is longitudinally uniform in the lower thermosphere, whereas it is more localized in the upper thermosphere (Figure \ref{fig:f6_wind_gw_pres_lon}). The propagation and dissipation of additional waves increase longitudinal variability of the zonal wind in the thermosphere up to $\sim$150 km. Above this altitude, the increase in variability is much less significant. This pattern can be explained as follows: As altitude increases, ion-neutral collisional interactions (i.e., ion drag) and the resulting  Joule heating become more important drivers of neutral variability. Despite the fact that gravity wave drag variability (including effects from non-tropospheric sources) increases sharply above 104 km and peaks in the upper thermosphere, these processes eventually overshadow (i.e., ion-neutral coupling) the contribution of waves in producing variability.

\subsection{Local Time Variability}

\begin{figure}[t!]
    \centering
    \hspace*{-1.5cm}
    \includegraphics[width=1.25\textwidth]{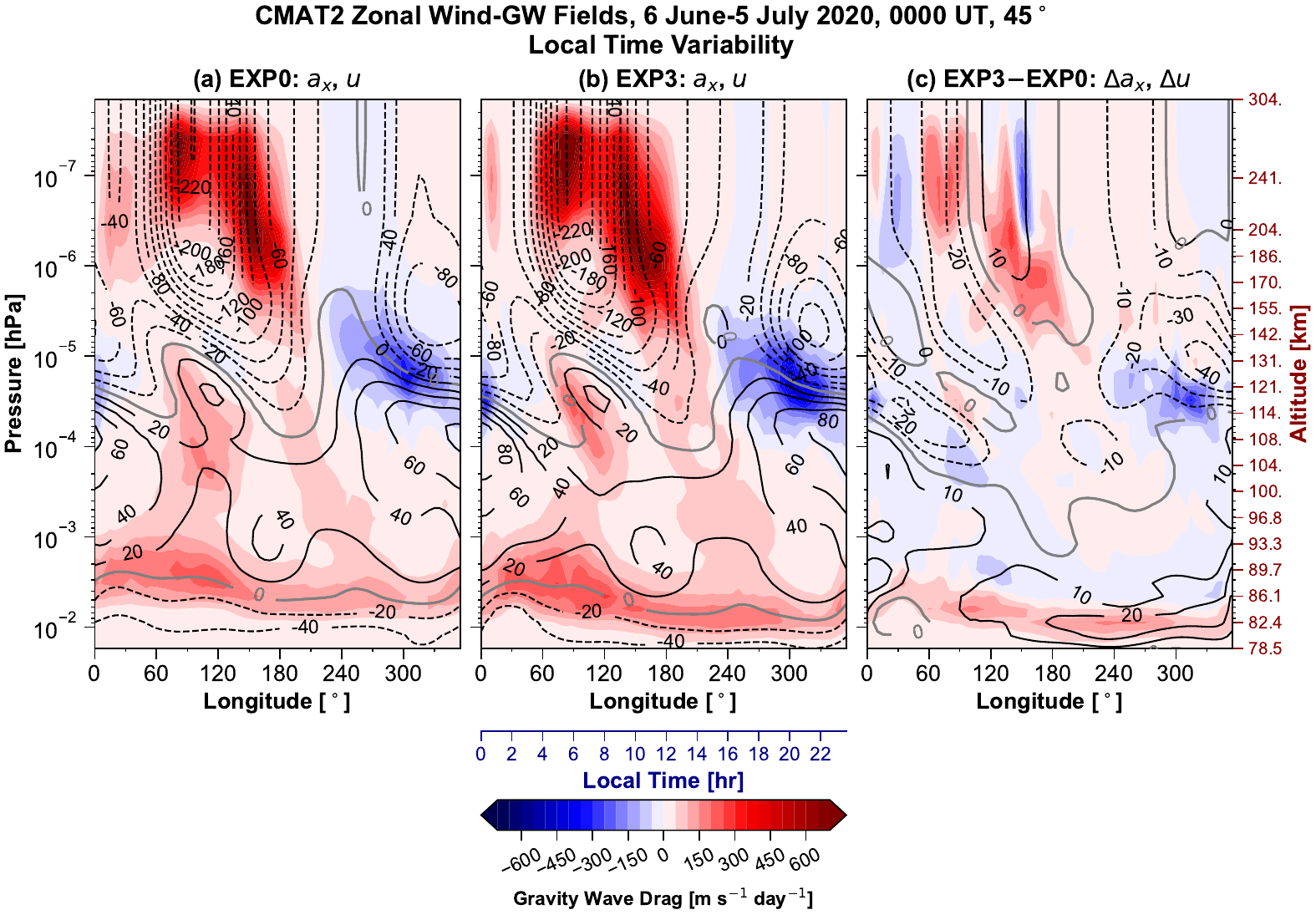}
    \caption{Same as Figure \ref{fig:f6_wind_gw_pres_lon} but for time-averaged at 0000 UT, showing time-average local time variability of the zonal wind and zonal gravity wave drag}
    \label{fig:f7_wind_gw_pres_lon_0000UT_mean}
\end{figure}

Since the background atmospheric fields (e.g., $T,\rho, u$) that shape gravity wave damping vary with local time, the upper atmospheric gravity wave drag is expected to vary with time as well.  \Fig \ref{fig:f7_wind_gw_pres_lon_0000UT_mean} shows the mean zonal wind and zonal gravity wave drag in the same format as in Figure  \ref{fig:f6_wind_gw_pres_lon} but time-averaged at 0000 UT. This  presents essentially the local time variations of the averaged winds and wave drag because longitude variation at a fixed UT can be viewed as a local time variation (shown in blue $x$-axis). During this season, the dawn and dusk are at 4:15 and 19:45 local time, respectively. Hence, the daytime lasts approximately between  4:15 and 19:45 LT, while the rest of the day corresponds to  the nighttime.

Studying local time variations provides further insight into  wind and wave variability. Thermospheric zonal winds vary strongly as a function of local time, with faster daytime westward winds than nighttime winds. At northern summer midlatitudes,  the pressure gradient  and  Coriolis forces are the dominant drivers of the neutral horizontal winds with ion drag, and gravity waves contributing significantly to the momentum balance. In the lower thermosphere, semidiurnal migrating tidal activity (SW2) maximizes, as evidenced by  variations in gravity wave drag and  zonal winds around 120 km. 

Mesospheric gravity wave effects have small local time variations, whereas thermospheric gravity wave effects vary greatly with local time. However, the lower thermospheric and upper thermospheric gravity wave activity exhibit very different local time behavior. In the upper thermosphere, gravity wave effects are much stronger during the day than at night. In contrast, in the lower thermosphere ($z\sim$ 100--140 km), nighttime gravity wave effects are stronger than the daytime ones. In the lower thermosphere, the modulation of the gravity waves by the SW2 is also discernible. In addition to the obvious difference  in the upper atmospheric gravity wave activity between day and night, we can also note a clear difference between dawn and dusk.  The upper thermospheric gravity wave drag is greater at dawn than at dusk, while the lower thermospheric gravity wave drag is greater at dusk than at dawn. 

The difference figure  in panel c demonstrates how the local time variations of the mean zonal winds and gravity wave drag change with respect to the benchmark run.  The mesospheric winds in the reversal region during daytime are more eastward (up to 20 \wind), owing to the additional eastward gravity wave drag produced by additional waves. In the lower thermosphere, the nighttime enhancement of the westward gravity wave drag  contributes to the westward wind change at night. However, in the upper thermosphere, the direction of relative wind change (i.e.,  eastward or westward with respect to the benchmark case) does not always coincide with the change in the local time behavior of gravity wave effects. 
During pre-noon (10-12 LT), the secondary waves generally enhance the upper thermospheric eastward gravity waves drag by up to 200\drag, which accelerates the wind only by 10\wind eastward. It is noteworthy that the upper thermospheric strong eastward change of the gravity wave drag after dawn coincides with a westward change of more that --20\wind of the thermospheric wind, suggesting that modification of the pressure gradient force and the ion drag could play a stronger role in controlling the changes in the neutral wind flow.  Another perspective  provided by this analysis is that the secondary wave can propagate to higher altitudes after dawn before noon than at night. 


\section{Comparison of CMAT2 with ICON Horizontal Winds}
\label{sec:cmat2-icon}

\begin{figure}[t!]
    \centering
    \hspace*{-1.3cm}
    \includegraphics[width=1.2\linewidth]{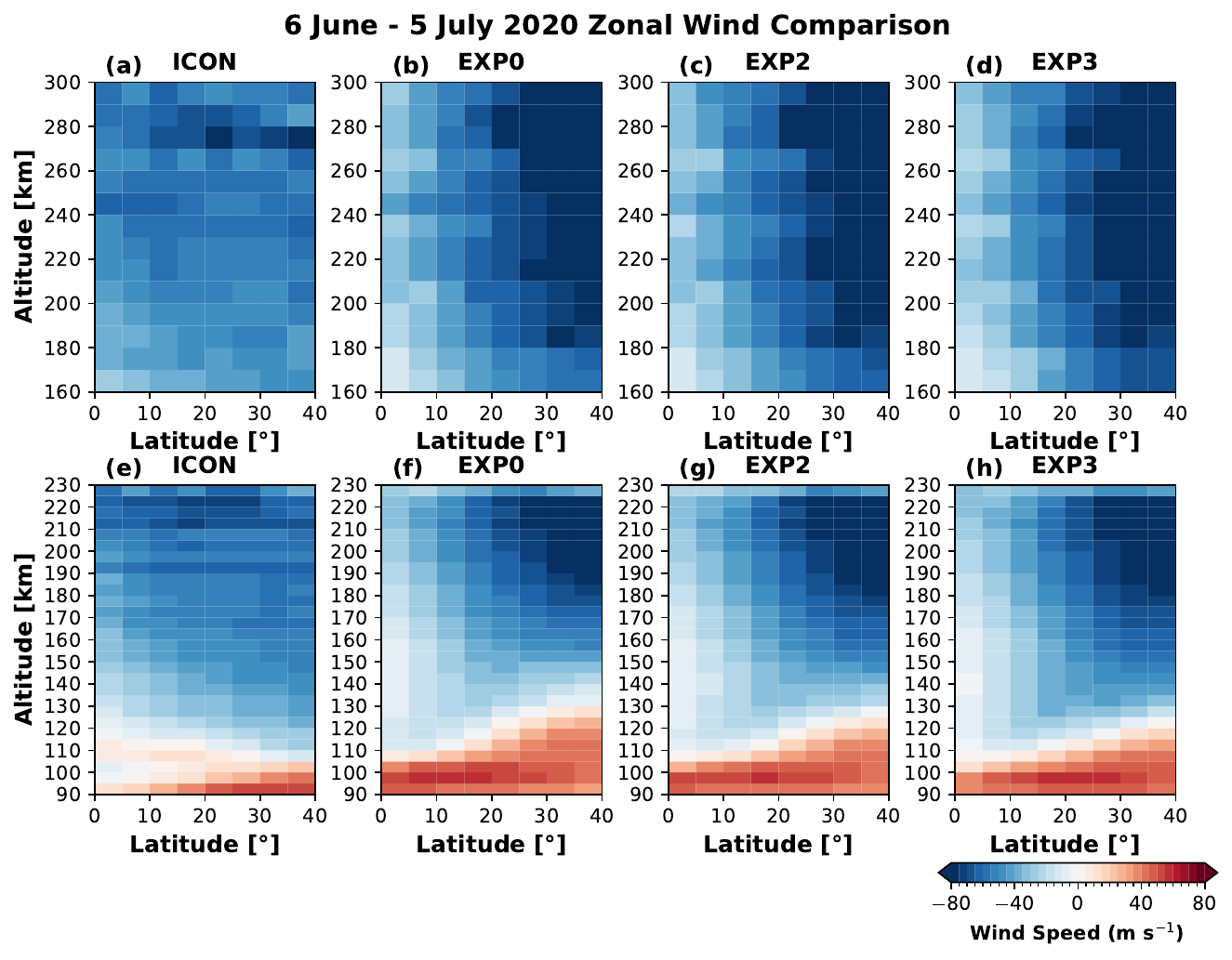}
    \caption{Altitude-latitude cross-sections of the mean zonal  wind  for June 6--July 5, 2020. The comparison includes daytime ICON-MIGHTI observations (panels a and e) compared against CMAT2 model configurations EXP0 (panels b and f), EXP2 (panels c and g), and EXP3 (panels d and h) across redline (160-300 km, upper row) and greenline (90--230 km, lower row) altitude ranges.}
    \label{fig:icon-cmat2-zonal}
\end{figure}

\begin{figure}[t!]
    \centering
    \hspace*{-1.3cm}
    \includegraphics[width=1.2\linewidth]{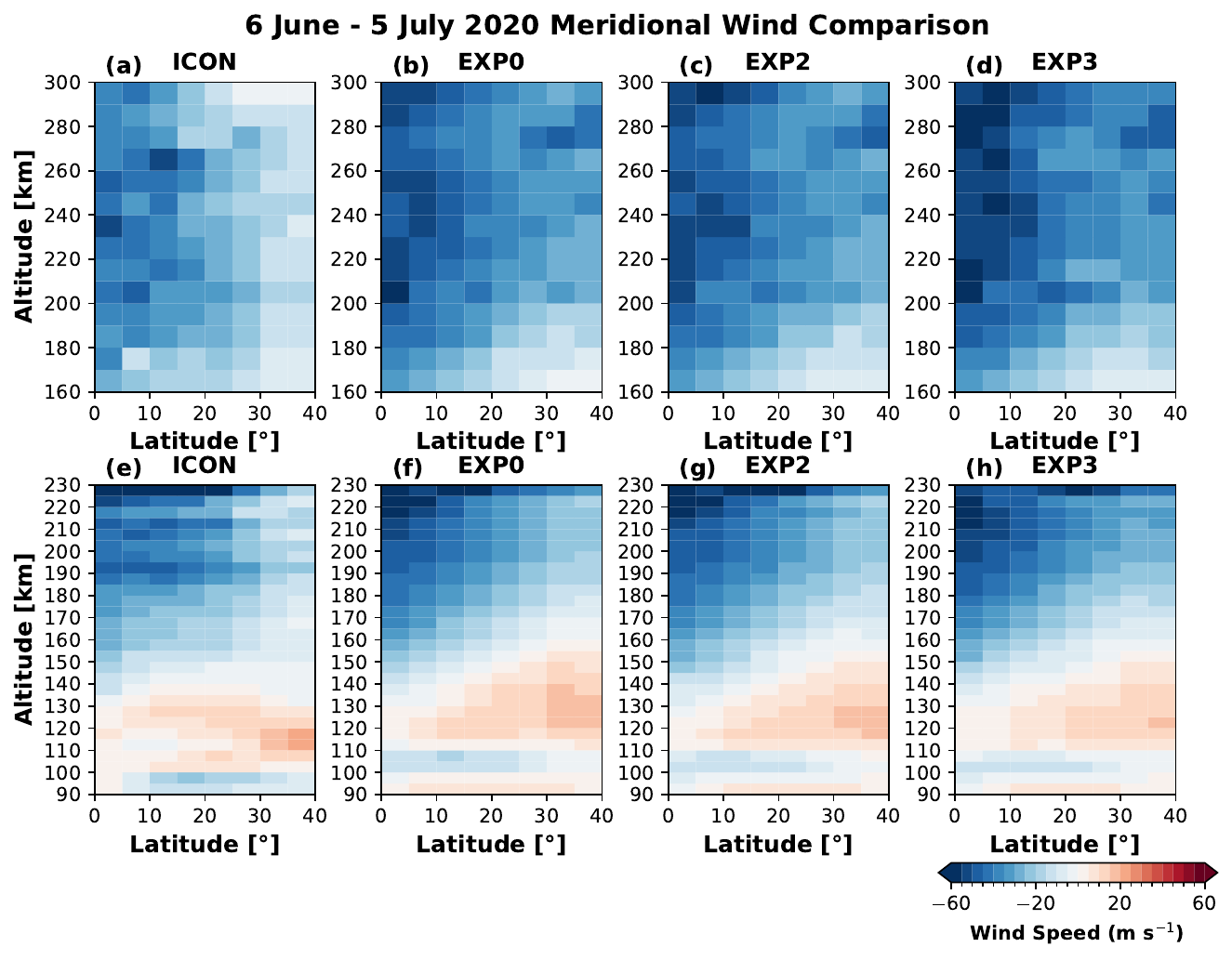}
    \caption{Same as in Figure \ref{fig:icon-cmat2-zonal} but for the meridional wind.}
    \label{fig:icon-cmat2-meridional}
\end{figure}

We compared three CMAT2 model configurations (EXP0: primary waves only; EXP2: localized sources at 50 km; EXP3: uniformly distributed sources) with daytime ICON/MIGHTI horizontal wind measurements from June--July 2020. \Figs \ref{fig:icon-cmat2-zonal} and \ref{fig:icon-cmat2-meridional} show that all model versions capture the fundamental vertical structure of thermospheric winds, particularly the transition from eastward to westward zonal flows with height. However, the model simulations show an eastward bias in the lower thermosphere and place the wind reversal at higher altitudes ($\sim 120$ km) than observed. Meridional winds display more complex patterns with weaker magnitudes, with observations showing northward flows around 100--130 km that are partially reproduced in the simulations, although with differences in spatial distribution. 

\begin{figure}[t!]
    \centering
    \hspace*{-1cm}
    \includegraphics[width=1.1\linewidth]{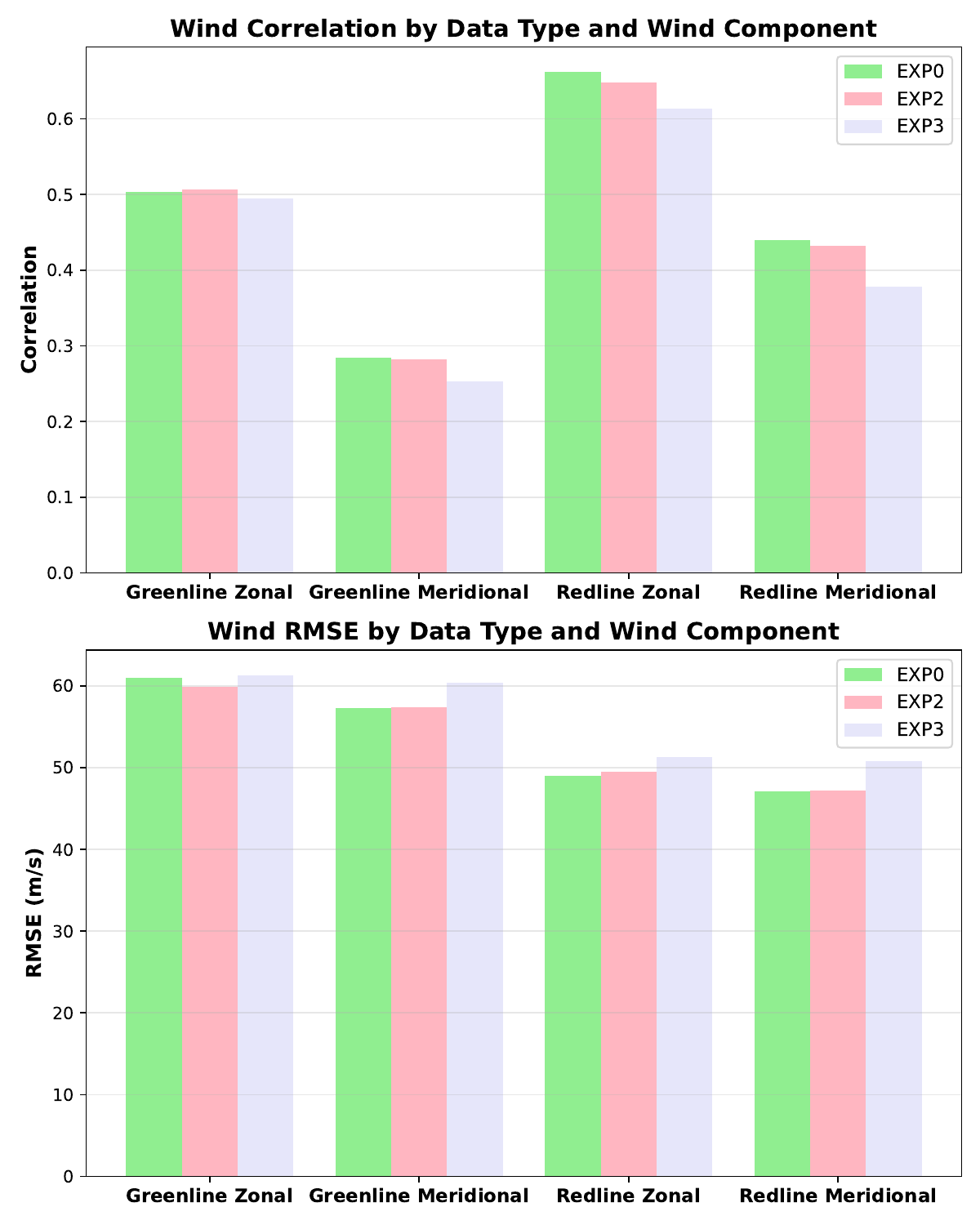}
    \caption{A statistical comparison between the CMAT-2 general circulation model simulations (EXP0, EXP2, EXP3) and daytime ICON-MIGHTI satellite observations of thermospheric winds. The top panel shows correlation coefficients and the bottom panel has root mean square error for both model configurations across different wind components (zonal and meridional) and emission types (greenline and redline).}
    \label{fig:validation-summary}
\end{figure}

As shown in  \Fig \ref{fig:validation-summary}, statistical comparisons reveal that the model performs better for zonal winds (correlation 0.5 - 0.65) than for meridional winds (correlation 0.25--0.45). Additionally the model shows a higher correlation with observations in the upper thermosphere (red line emissions, 160--300 km) than in the lower thermosphere (green line, 90--230 km). RMSE values range from 45--60 \wind, with lower errors for the upper thermospheric measurements. The addition of extra-tropospheric wave sources in EXP2 and EXP3 does not improve- and even slightly degrades the statistical metrics compared to EXP0, despite producing qualitative improvements in wind structures in the wave breaking region (110-140 km). This suggests that either the effects of these waves are too spatially localized to affect global statistics, or that other physical processes unrelated to gravity waves dominate the remaining model-observation discrepancies.

\section{Summary \& Conclusions}
\label{sec:summary}
We have studied the relative dynamical roles of gravity waves of tropospheric and non-tropospheric origin, using the Coupled Middle Atmosphere Thermosphere-2 general circulation model, which extends from the tropopause to the upper thermosphere. The model includes the whole atmosphere gravity wave parameterization, as described by \citeA{Yigit.etal2008_ParameterizationEffects}, with wave sources incorporated using the framework developed by \citeA{Medvedev.etal2023_DynamicalImportance}. Since the exact mechanisms of wave generation are not well understood, and therefore not parameterized, the sources are expressed in terms of multiples of the tropospheric momentum forcing, which is required to produce the incident primary momentum flux spectrum at the lower boundary. We performed GCM simulations with primary waves of tropospheric origin and with extra-tropospheric sources (i.e., localized sources at 50 and 90 km, and sources distributed over all heights). The main inferences of this study are the following.
\begin{enumerate}
\item  By adding momentum to the harmonics propagating from below, extra-tropospheric sources contribute to the effects produced by gravity waves in the upper atmosphere. 
\item Differences in the simulated  zonally and time-averaged temperature and horizontal wind fields demonstrate that sources uniformly distributed over all heights above the tropopause produce the greatest change in the mesosphere and thermosphere, where they can modify the mean zonal winds by up to $\pm 30$ m~s$^{-1}$. The  changes produced by localized sources are considerably smaller, even though their forcing was assumed to be ten times stronger than in the troposphere. Changes induced by sources centered at 50 km are about three times smaller than those induced by sources distributed at all heights. When gravity wave sources are localized around 90 km, the thermospheric effects are negligible. The main changes occur in regions where the wind reverses in the upper mesosphere and lower thermosphere.
\item Although the sources were assumed to always add the momentum to the incident waves originating in the troposphere, they do not necessarily enhance the impact of the latter, as seen in the midlatitude mesosphere and thermosphere. Occasionally, the waves from additional sources can weaken the original effects by creating drag in the opposite direction. 
The heating/cooling rates in simulations with non-tropospheric sources can be by up to 40 K~day$^{-1}$ larger than in the simulations with only tropospheric sources.
\item Longitudinal variability of both the zonal wind and drag increases with altitude. This is because the gravity wave drag is longitudinally uniform in the lower thermosphere but more localized in the upper thermosphere. Including sources distributed over all heights  increases the longitudinal variability of the zonal wind in the thermosphere up to $\sim$150 km. Above this height, the variability increases to a much lesser extent, primarily due to the competition with other forces in the thermosphere.
\item  Propagation and dissipation of gravity waves in the thermosphere depend on local time due to variability in the background atmosphere and dissipative processes. Local time variations of gravity wave effects differ between the lower (100--140 km) and upper thermosphere. Wave effects in the upper thermosphere are much stronger during the day than at night. Conversely, nighttime gravity wave effects are stronger than daytime ones in the lower thermosphere.  The modulation of  gravity waves by the semidiurnal migrating tide is clearly discernible in the lower thermosphere. 
\item  A comparison of ICON-MIGHTI observations and CMAT2 results shows that CMAT2 reproduces the basic structure of thermospheric winds. However, it performs better with  zonal winds than with meridional winds. The model exhibits an eastward bias in the lower thermosphere and places wind reversals higher than observed. Adding extra-tropospheric wave sources (EXP2 and EXP3) modifies wind structures in the wave-breaking regions, but it does not improve the global statistical comparison. 
\end{enumerate}


Taking a gravity wave spectrum equal in strength to that produced by tropospheric waves and imposing it at every altitude in the middle atmosphere is admittedly an extreme case of secondary waves. Currently there is no strong evidence for such persistent and extensive wave generation in the middle atmosphere. In the absence of such evidence, this study suggests that additional wave generation in the middle atmosphere would have relatively minor effects on the circulation in the thermosphere.

If the localized sources in the middle atmosphere are associated with secondary wave sources, then assigning them ten times stronger than in the troposphere is a great exaggeration. What is well understood is that the breaking processes that generate secondaries also dissipate energy and transfer momentum to the mean flow (wave drag). Therefore, the secondary sources must be weaker than the primary waves that generated them. Also significant gravity wave sources need to be forced by large vertical motions, and these tend to be suppressed in the middle atmosphere.

Since the hypothetical middle atmosphere wave sources considered in this study mainly add to the dynamical effects of primary waves, one might anticipate that if new statistics on middle atmosphere wave generation do become available, these could be accommodated by retuning existing wave drag parameterizations. 

In this study, we have assumed that the spectral shape and the characteristic horizontal wavelengths of the  waves generated by tropospheric and non-tropospheric sources are similar. However, this is not necessarily true for different sources. Future studies can explore different spectral shapes and/or horizontal scales of waves. 

We assumed in our simulations that the extra-tropospheric sources always generate waves in sync with the incident spectrum. In other words, they always enhance  the incident harmonics. However, this  is highly unlikely to occur in  reality, because phase shifts between the tropospheric and non-tropospheric waves could vary from increasing to completely canceling out the tropospheric waves. Furthermore, it is an exaggeration to assume that sources are monotonically distributed at all heights.  Therefore, these simulations should be considered as an upper estimate of potentially missing gravity wave sources. Finally, given the lack of knowledge regarding mechanisms of secondary wave generation, the localized sources (EXP1 and EXP2) could serve as proxies.

  \begin{acronyms}
  \acro{CMAT2-GCM}
  Coupled Middle Atmosphere Thermosphere-2 General Circulation Model
  \acro{ICON} Ionospheric Connection Explorer
  \acro{GWs} Gravity Waves
  \acro{MIGHTI} Michelson Interferometer for Global High-resolution Thermospheric Imaging
  \acro{NCEP} National Centers for Environmental Prediction
\end{acronyms}

\section*{Data Availability Statement}
The data on which this article is based are available in the following datasets:

ICON/MIGHTI horizontal wind data (version 5) \cite{Immel.etal2018_IonosphericConnection} 
are available via ICON data center
URL: \url{https://icon.ssl.berkeley.edu/Data}.
The data for the coulumn model and the GCM presented in this paper \cite{Yigit.etal2025_DatasetImpact} are available via Zenodo: \url{https://zenodo.org/records/15476686}.

\acknowledgments
EY and ALSG were supported by NASA (Grant 80NSSC22K0016) and National Science Foundation (Grant AGS 2330046).  


\clearpage
\newpage
\section*{Figures}
%
%

%
%

%
%

%
%

%
%

%
%

%
%

%
%

%
%

%
%

%
%

%
%

\newpage
\end{document}